# HIGH ENERGY LHC WORKSHOP (MALTA, OCTOBER 2010) SUMMARY OF SESSION 3 ON "SYNCHROTRON RADIATION AND BEAM DYNAMICS"

V. Shiltsev (Fermilab) and E. Métral (CERN)

*Abstract*

Below we summarize presentations, discussions and general conclusions of the Workshop session on "Beam Dynamics Issues". Major subjects include effects due to synchrotron radiation (SR), cryogenic loads, electron cloud, impedances, intra-beam scattering (IBS) and beam-beam interactions.

## INTRODUCTION

The Charge to the Workshop is to "… take a first look at a higher-energy LHC (HE-LHC) with about 16.5 TeV beam energy and 20-T dipole magnets", therefore, in the AM session Friday October 15, we have concentrated our efforts onto understanding and evaluation of the potential issues with beam dynamics in HE-LHC and identification of the topics for future, more technical studies.

There were seven presentations on the subject: "Heat load and cryogenics" by Dimitri DELIKARIS (CERN) [1]; "Requirements from the vacuum system" by Jose Miguel JIMENEZ (CERN) [2]; "Beam screen issues" by Elias METRAL (CERN) [3]; "IBS and cooling at RHIC and HE-LHC active emittance control" by Wolfram FISCHER (BNL) [4]; "Modeling IBS and cooling" by Oliver BOINE-FRANKENHEIM (GSI) [5]; "SR damping, IBS, and beam-beam simulations" by Alexander VALISHEV (FNAL, presented by V. Shiltsev) [6]; "SR and beam-beam simulations" by Kazuhito OHMI (KEK) [7].

## CRYOGENICS, VACUUM LOAD AND BEAM SCREEN

The HE-LHC will be the first hadron machine dominated by synchrotron radiation. Compared to design LHC parameters, it will see 17-fold increase of the SR power from 0.33 to 5.7 W/m. The analysis performed in Ref. [1] shows that the total heat load on the beam screen (SR + image current heating + rest) will be about 10 W/m and suggests that the optimal temperature of the beam screen is in the range 40-60 K (vs. 4.5-20 K now). The optimal temperature of the magnet cold mass is 2K as it allows some ~ 2 T higher peak dipole filed (and thus, more than 10% higher energy) and also greatly helps to assure field stability in the magnet. Equivalent total HE-LHC cryo capacity is about what LHC has now, but how much of that could be refurbished in ~ 2030 (after > 20 years of operation) is now clear yet.

It was noted in Ref. [2] that the resistivity of the 40-60 K beam screen is ~ 5.5 higher than in the LHC, and in addition, higher dipole magnetic field will cause an additional factor of ~ 2 increase due to the magneto-resistance effect in the higher (20 T) field [3].

It was also found that anomalous skin effect will be negligible [3]. In total, the resistive wall (RW) impedance of the beam screen which scales as $\rho^{1/2}$ will be a factor 3.3 higher than in the LHC but probably that is not of great concern (from the point of view of the beam instabilities) because the beam energy will be higher by a factor of 2.4 at "flat top" or 2-3 at the injection (if a higher energy injector will be built). The discussions in the group ended up in an overall conclusion that instabilities should not be a major issue in the HE-LHC but further considerations will be needed. Among various ideas to reduce instabilities we discussed a possibility of a superconductive HTS coating – which was found to be not appropriate as that will keep the magnetic flux frozen and forbid ramping of the machine – and use of Al screen to reduce impedance and magneto-resistance – that option is not too advantageous either because of higher e-cloud yield.

What was found of significant practical concern is the beam-induced pressure rise in HE-LHC (see Ref. [2]). The flux and energy of the SR photons radiated inside the beam screen will be significantly higher than those in the LHC that will lead to about 74 (!)-fold increase in the beam-induced pressure rise. So far, no single solution of the problem was found, so a number of measures were offered to keep the problem under control: a) Increase pumping speed with larger area of slots in the beam screen (now ~ 4%, can possibly be doubled); b) Use TiN or amorphous-C coating in cold sectors to control electron cloud formation; c) Consider use of clearing electrodes (say, + 500V strip all along the beam pipe) or solenoids; d) NEG coating in warm sectors (where it is possible to bake the pipe to activate the coating); e) One can also count on the vacuum cleaning by SR and e- bombardment and beam scrubbing (by losses) – that will take time, and may force to start operation with a low number of protons per bunch. The overall conclusion on the issue was that at the moment, the vacuum does not look as the HE-LHC showstopper, but that is something definitely to be concerned of, and a more detail study of the issue will be required, based on the LHC experience.

# SYNCHROTRON RADIATION DAMPING EFFECTS, INTRA-BEAM SCATTERING AND BEAM-BEAM EFFECTS

Contrary to other high energy hadron colliders, in the HE-LHC the SR emittance damping times - of about 1 hour (long.) and 2 hours (transv.) – will be much shorter than the IBS growth times (> 50h), thus, the SR will dominate the luminosity dynamics unless beam-beam or other effects will be stronger. During the presentations [4-7] and in the following discussions it has been shown that the SR damping/fluctuations and their effects on the beam dynamics are well understood [4,6,7]; the IBS theory, and proven models and simulation codes are available [4,5]; the initial HE-LHC luminosity integral estimates of ~ 0.8 fb$^{-1}$/day are correct and confirmed by others [4,6,7].

The understanding of the beam-beam effects is somewhat poorer and the predictive power of modern beam-beam modeling tools is limited. The design beam-beam parameter in the HE-LHC is not outstandingly high compared to other machines and the LHC start up conditions (see Fig.1).

It was noted that experience from the LHC operation will be quite important to make predictions for the HE-LHC. It will tell which kind of bean dynamics phenomena sets the most stringent limits on the luminosity performance: a) Instabilities; b) Head-on or/and long range beam-beam effects; c) Intolerable beam losses; d) Emittance blowups; e) Beam luminosity/lifetime; f) Collimation system (in)efficiency; g) External noises, drifts; h) Some other effects or combination of the above mentioned effects. (At the current stage of 1% of the design luminosity – it seems to be too early to draw conclusions and make strong recommendations for the HE-LHC on the basis of the LHC performance).

It was brought up in the discussions that on one hand, in the HE-LHC: the luminosity burn up and the SR damping will dominate the luminosity evolution and daily integral; the IBS does not matter to a ~ 1% level; the beam-beam effects do not matter ~ 10% level; while on the other hand, there are several interesting questions to answer: a) Does the SR damping/cooling help to increase beam-beam limit?; b) If "yes", then by how much? Can one count on the parameter *ξ > 0.01/IP*)?; c) Can even faster beam cooling help further? E.g. the so called Optical Stochastic Cooling [9] or coherent electron cooling [10] can give extra < 1 hour of the emittance cooling decrement reduction; d) Is some kind of beam heating (controlled emittance blow up) needed to stay at the beam-beam limit or the beam-beam induced emittance blow up can stabilize itself (e.g. in Tevatron b-b emittance blowup is much faster than 1 hour)? [11]; e) How effective might be various compensation schemes: e.g. electron lenses [12], current carrying wires [13], "crab waist" collision scheme with flat beams [14]?; f) How serious are the concerns of coherent beam-beam instabilities, and in particular, multi-bunch beam-beam phenomena?

Although at present, synchrotron radiation, IBS and beam-beam effects do not seem to pose major concerns, the questions raised above are better be carefully studied.

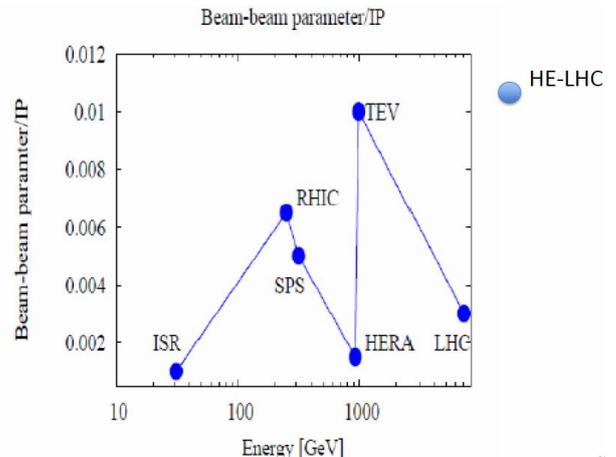

Figure 1: Beam-beam parameter in the hadron colliders, from Ref. [8].